\documentclass[11pt]{article}
\usepackage{amsmath,amsfonts}
\def \beq  {\begin{equation}}
\def \eeq  {\end{equation}}
\def \beqar {\begin{eqnarray}}
\def \eeqar {\end{eqnarray}}
\def\sqr#1#2{{\vcenter{\vbox{\hrule height.#2pt
\hbox{\vrule width.#2pt height#1pt \kern#1pt
\vrule width.#2pt}\hrule height.#2pt}}}}

\def\la {{\langle}}
\def\ra {{\rangle}}

\def\Tr {{\rm Tr}}
\def \tr {{\rm tr}}

\def\del {\partial}

\def\D {{\cal D}}
\def\bz {{\bar{z}}}

\def\half{\textstyle{1\over 2}}

\begin{document}
%%%%%%%%%%%%%%%%%%%%%%%%%%%%%%%%%%%%%%%%%%%%%%%
%\fontfamily{pnb}\fontsize{12pt}{16pt}\selectfont
%\fontfamily{pzc}\fontsize{14pt}{16pt}\selectfont
%\fontfamily{pbk}\fontsize{12pt}{16pt}\selectfont
%\fontfamily{cmr}\fontsize{11pt}{15pt}\selectfont
%\fontfamily{phv}\fontshape{ro}\fontsize{11pt}{16pt}\selectfont
%\fontfamily{ptm}\fontseries{m}\fontshape{r}\fontsize{12pt}{16pt}\selectfont
%\fontfamily{pnc}\fontseries{m}\fontshape{r}\fontsize{11pt}{15pt}\selectfont
%\usefont{T1}{phv}{m}{it}
%%%%%%%%%%%%%%%%%%%%%%%%%%%%%%%%%%%%%%%%%%%%%%%
\def \CMP {{ Commun. Math. Phys.}}
\def \PRL {{ Phys. Rev. Lett.}}
\def \PL {{Phys. Lett.}}
\def \NPBProc {{ Nucl. Phys. B (Proc. Suppl.)}}
\def \NP {{ Nucl. Phys.}}
\def \RMP {{ Rev. Mod. Phys.}}
\def \JGP {{ J. Geom. Phys.}}
\def \CQG {{ Class. Quant. Grav.}}
\def \MPL {{Mod. Phys. Lett.}}
\def \IJMP {{ Int. J. Mod. Phys.}}
\def \JHEP {{ JHEP}}
\def \PR {{Phys. Rev.}}
\def \JMP {{J. Math. Phys.}}
%%%%%%%%%%%%%%%%%%%%%%%%%%%%%%%%%%%%%%%%%%%%%%%
%%%%%%%%%%%%%%%%%%%%%%%%%%%%%%%%%%%%%%%%%%%%%%%
\begin{titlepage}
\null\vspace{-62pt} \pagestyle{empty}
\begin{center}
\rightline{CCNY-HEP-06/5}
\rightline{April 2006}
\vspace{1truein} {\Large\bfseries
The Chern-Simons One-form and Gravity on a}\\
\vskip .2in
{\Large\bfseries  Fuzzy Space}\\
\vskip .2in\noindent

%%%%%%%%%%%%%%%%%%%%%%%%%%%%%%%%%%%%%%%%%%%%%%%%%
\vspace{.5in}
{\bf\large V. P. NAIR}\\
\vspace{.15in}{\itshape Physics Department\\
City College of the CUNY\\  
New York, NY 10031}\\
E-mail:
\fontfamily{cmtt}\fontsize{11pt}{15pt}\selectfont {vpn@sci.ccny.cuny.edu}

\vspace{.4in}
%\vspace{1.5in}%\vspace{0.3in}
%%%%%%%%%%%%%%%%%%%%%%%%%%%%%%%%%%%%%%%%%%%%%%%%%%%%%%%%%%%%
\centerline{\large\bf Abstract}
\end{center}
The one-dimensional ${\cal N}\times {\cal N}$-matrix Chern-Simons action is given,
for large ${\cal N}$ and for slowly varying fields, by
the $(2k+1)$-dimensional Chern-Simons action $S_{CS}$,
where the gauge fields in 
$S_{CS}$ parametrize the different ways in which the large ${\cal N}$ limit
can be taken. Since some of these gauge fields correspond to the
isometries of the space, we argue that gravity on fuzzy spaces can be described by the
one-dimensional matrix Chern-Simons action at finite ${\cal N}$
and by the higher dimensional Chern-Simons
action when the fuzzy space is approximated by a continuous manifold.

\end{titlepage}
%%%%%%%%%%%%%%%%%%%%%%%%%%%%%%%%%%%%%%%%%%%%%%%%%%%%%%
\pagestyle{plain} \setcounter{page}{2}

\section{Introduction}

Noncommutative and fuzzy spaces have been interesting objects of research for a long 
time \cite{general}. Such spaces can be realized as solutions, for example, branes, sometimes with specific background fields, in string theory and in the matrix version of $M$-theory \cite{general2}.
Gauge theories on such spaces are interesting since they can describe fluctuations of the brane solutions. Fuzzy spaces are noncommutative spaces which can be described by finite dimensional matrices.
Examples of many fuzzy spaces, such as $S^2_F$, ${\bf CP}^k_F$, etc., have been constructed. When the dimension of the matrices which represent the coordinates becomes large, these spaces tend to their smooth counterparts $S^2$, ${\bf CP}^k$, etc.
The finite dimensionality of the matrices means that there are only a finite number of modes for any fields defined on a fuzzy space. Therefore we may regard such theories as finite-mode approximations to the usual field theories on a smooth (and commutative) manifold
\cite{bal1}. It is thus an alternative method of regularization, analogous to the lattice regularization. It is also possible to preserve more symmetries in a fuzzification, compared to latticization
\footnote{There may be other advantages to fuzzification as well. For example, it has been argued that inherent nonlocality of fuzzy spaces can be used to evade the fermion doubling problem of the lattice formulations in a way that still preserves most of the desirable symmetries \cite{bal2}.}.
The matrix description of branes can also be viewed as a regularization of the brane itself, and in this sense, it has become part of the standard repertoire of methods for analyzing branes \cite{BMN}.
While, for most theories, a new regulator may not mean very much, gravity is one case where fuzzy spaces could do significantly better, due to the possibility that fuzzification can preserve symmetries. Naturally there have been many investigations into gravity on fuzzy spaces, and more generally, on noncommutative spaces \cite{ncgrav, nair1}. Clearly, it is important to explore gravity on fuzzy spaces.

One can even go a bit further in these considerations, raising the question whether it is possible to describe the world by a finite-dimensional Hilbert space. Since there are many spaces, such as de Sitter space, which have finite entropy, a finite-dimensional Hilbert space describing all physical degrees of freedom is worthy of consideration, not just as a regulator, but as a basic premise for physical theories.

Fuzzy spaces also have an obvious connection to the lowest Landau level of a quantum Hall system \cite{KNR}. This is most easily seen with a specific example, say, ${\bf CP}^k =  SU(k+1)/U(k)$.
The isometries of this space are given by $SU(k+1)$, with $U(k)$ as the local isotropy group, the analog of the Lorentz group  for Minkowski space. The Riemann curvature has values in the Lie algebra of $U(k)\sim SU(k)\times U(1)$, and has constant components in the tangent frame basis. To formulate the Landau problem, consider ${\bf CP}^k$ with an additional constant ``magnetic field'' proportional to, say, the $U(1)$-component of the Riemann tensor.
The lowest Landau level can then be characterized by a basis of wave functions which are, evidently, sections of a $U(1)$-bundle on ${\bf CP}^k$. The set of such lowest Landau level wave functions form a finite dimensional Hilbert space ${\cal H}_N$. Observables of the theory, for all dynamics projected to the lowest Landau level, are then arbitrary hermitian
$(N\times N)$-matrices. The algebra of such observables is obviously $Mat_N$, the
matrix algebra of $(N\times N)$-matrices. With a suitable choice of a Laplacian $\Delta_N$, the triplet $({\cal H}_N, Mat_N, \Delta_N )$ can then be taken as the definition of ${\bf CP}^k_F$, the fuzzy version of ${\bf CP}^k$.
It is easy to show, by standard techniques of semiclassical analysis, that $Mat_N$ tends
to the algebra of functions on ${\bf CP}^k$ as $N$ becomes large.

The quantum Hall system thus gives us a model and a physical context to think about gravity on fuzzy spaces. (For a general discussion of various aspects of the quantum Hall
system in higher dimensions, see references \cite{ZH, KN, other}.)
As for the dynamics, one can consider a Hall droplet of fermions for which the bulk and boundary dynamics have been analyzed in some detail \cite{KN}. In what follows, it is the bulk dynamics which will be relevant. By virtue of the exclusion principle and the related incompressibility of the droplet, the bulk dynamics will be trivial, unless one introduces fluctuations in the gauge fields (or ``magnetic fields'').
Local changes in gauge fields can change the local density of states leading to rearrangements of fermions in the droplet, including changes in ddensity. Since such gauge fields can take values in $\underline{SU(k+1)}$ for ${\bf CP}^k$, they can be interpreted as gauging the isometries of ${\bf CP}^k$. Naturally, this suggests a way of introducing gravitational fields on a fuzzy space \cite{nair2}.\footnote{It is also suggestive that
quantum Hall droplets appear in the dual field theories for many gravitational backgrounds
\cite{beren}.}

There is one more important insight which can be elicited from this mapping of the fuzzy gravity problem to the quantum Hall system.
Imagine we have a droplet of fermions of finite size in the lowest Landau level. The droplet may be characterized by a density matrix $\rho_0$. The general time evolution of the droplet is then given by a unitary tranformation $\rho_0 \rightarrow U \rho_0 U^\dagger$, obeying the standard quantum Liouville equation,
\beq
i {\del \rho \over \del t} = [K, \rho ]
\label{intro1}
\eeq
where $K$ is the Hamiltonian. The action which leads to this equation is
\beqar
S&=& \int dt~ \Tr \rho_0 \bigl( i U^\dagger {\dot U} - U^\dagger K U \bigr)
= i \int dt~ \Tr \rho_0 \bigl( U^\dagger D_0 U\bigr)\label{intro2}\\
D_0 &=& {\del \over \del t} + A_0, \hskip .3in A_0 = iK\nonumber
\eeqar
In reinterpreting the Hilbert space as defining a fuzzy space, the states correspond to points.
This suggests that the dynamics of space itself (gravity) should be the same as the evolution of states in the quantum Hall image of it, and, hence, it is given by the action (\ref{intro2}).
This what the quantum Hall analogy gives us: The action (\ref{intro2}), with some qualifications elaborated on below, should be the action for fuzzy gravity. 
{\it Our basic principle is that the same equation for evolution should hold for the states corresponding to
space as for states corresponding to matter.}
We shall now explore this in more detail, but need
some technical results.

\section{The one-dimensional Chern-Simons action}

We will formulate the calculation of the effective action, which is the large $n$ limit 
of an action like (\ref{intro2}), using ${\bf CP}^k= SU(k+1)/U(k)$ as the phase
space. It is possible to choose other spaces for this purpose as well. Th final result we are aiming for will be independent of the specific choice.
As mentioned before, constant magnetic fields on ${\bf CP}^k$ take values in 
$\underline{U(k)}$, the Lie algebra of $U(k)$. Wave functions can be given in terms of
a general $SU(k+1)$ element, say, a $(k+1)\times(k+1)$-matrix $g$.
We will denote by $t_a$, $a=1,2, \cdots , k^2+2k$, a set of hermitian matrices which form
a basis for the Lie algebra of $SU(k+1)$ in the fundamental representation, with
\beq
[ t_a, t_b ] = i f_{abc} t_c , \hskip 1in \Tr (t_a t_b )= \half \delta_{ab}
\label{1dcs1}
\eeq
$f_{abc}$ are the structure constants of $SU(k+1)$ in this basis.
(We take what is the conjugate of the fundamental representation in the 
conventional sense. For example, for $SU(3)$, our matrices are given by
$t_a = -\half \lambda_a^T$, where $\lambda_a$ are the standard Gell-Mann matrices 
and the superscript
$T$ denotes the transpose.)
We also define the differential operators $L_a$, $R_a$ on $g$
corresponding to the left and right translations as
\beq
L_a ~g = t_a ~g, \hskip 1in R_a ~g = g~t_a
\label{1dcs2}
\eeq
It is convenient to split these into the $R_{k^2+2k}$ which is the $U(1)$ generator
in $U(k)\subset SU(k+1)$, $R_j$, $j= 1, 2, \cdots, k^2-1$, which are $SU(k)$ generators
and $R_{\pm i}$, $i=1, 2, \cdots, k$ which are in the complement of
$\underline {U(k)}$ in the Lie algebra $\underline{SU(k+1)}$.

For our purpose, it is sufficient to consider a background magnetic field which is valued in
$\underline{U(1)}$. The fact that we have a $U(1)$ background field can be expressed 
on the wave functions as \cite{KN}
\beqar
R_j ~\Psi (g)&=&0, ~~~~~~~~~~~~~~~j=1,\cdots , k^2 -1 \nonumber \\
R_{k^2 + 2k} ~\Psi (g)&=& nk {1 \over {\sqrt{2k(k+1)}}}~\Psi (g)
\label{1dcs3}
\eeqar
where $n$ is an integer characterizing the strength of the field. 
For the lowest Landau level, we also have
\beq
R_{+i}  ~\Psi (g) =0
\label{1dcs4}
\eeq
The wave functions obeying the conditions
(\ref{1dcs3}) and (\ref{1dcs4}) are given by
\beq 
\Psi_m (g) = \sqrt{N}~ {\cal D}^{(n)}_{m,n}(g)\label{1dcs5}
\eeq
where the 
Wigner ${\cal D}$-functions of
$SU(k+1)$ are defined as
\beq
{\cal D}^{(n)}_{pq}(g) \equiv \la n, p \vert~ {\hat g}~ \vert n, q\ra
\label{1dcs6}
\eeq
They are the representatives of $g$ in the rank $n$ symmetric representation; the right
index on ${\cal D}^{(n)}_{m,n}$ occurring in (\ref{1dcs5}) is fixed to be $n$, ${\cal D}^{(n)}_{m,n} (g) =  \la n,m\vert
{\hat g}\vert n,n\ra$. Also, $N$ is the dimension of the representation, $N= (n+k)! /n! k!$.

The large $n$ expansion is most easily obtained using the symbol of a matrix defined by
\beq
({\hat A}) = A(g) =  \sum_{ms}\D^{(n)*}_{m, n}
(g) A_{ms} \D^{(n)}_{s, n}(g) = \la w\vert ~ {\hat g}^{-1} {\hat A}~ {\hat g}~\vert w\ra
\label{1dcs7}
\eeq
The symbol
of a matrix ${\hat A}$ coincides with the expectation value of ${\hat A}$
in the large $n$ limit.
Further,
the trace of a matrix 
is given in terms of its symbol by
\beq
\Tr~ {\hat A} = \sum_{m} A_{mm} = N~\int d\mu (g) ~A(g)
\label{1dcs8}
\eeq
where the integration is over the phase space, ${\bf CP}^k$ in our case.
The symbol corresponding to the product of two matrices is given by the star product
of the symbols of the matrices. It can be written as
\beq
({\hat A} {\hat  B})\equiv ({\hat A})* ({\hat B})
 = A B ~-{1\over n} {\hat R}_{-i} A ~{\hat R}_{+i} B ~+~{\cal O}(1/n^2)
 \label{1dcs9}
\eeq
We will not need the full expansion for the calculations presented here.
Equation (\ref{1dcs8}) and (\ref{1dcs9}) tell us that it is possible to replace the
trace of a product of matrices by the integral of the star product of the symbols of the matrices. In the star product, the leading term is just the product of the symbols, with corrections which involve derivatives, these being
subdominant at large $n$.

Fuzzy ${\bf CP}^k$ can also be described, before we take the large $n$ limit,
 in a way which reproduces its
embedding in ${\bf R}^{k^2 +2k}$ when $n$ becomes large.
In this case, ${\bf CP}^k_F$ is given by
$k^2 +2k$ hermitian matrices $X_a$ which are of dimension
$(N\times N)$ .
The embedding conditions are then given by \cite{bal3}
\beqar
X_a X_a &=& {nk(n+k+1) \over 2(k+1)}~\equiv C_n\nonumber\\
d_{abc} X_b X_c &=& (k-1) {(2n +k +1) \over 4 (k+1)}~ X_a ~\equiv \alpha_n ~X_a
\label{1dcs10}
\eeqar
where $C_n$  is the quadratic Casimir operator
and $\alpha_n$ is another invariant related to the properties of the $d_{abc}$-symbol.
These embedding equations are solved by
$X_a = T_a$, where $T_a$ are the $SU(k+1)$-generators in the symmetric representation
of rank $n$.
The generators $-iT_a$ also play the role of derivatives via their adjoint action.
On the fuzzy space, gauge fields are described by $D_a = -iT_a +A_a$, where $A_a$ is the gauge potential. It is still a matrix at this stage.
{\it A priori} there are $k^2 +2k$ spatial
components for the gauge potential, but there are restrictions on $D_a$ 
which ensure that there are
only $2k$ spatial components for the potentials, as expected for ${\bf CP}^k$. These conditions are
the gauged version of the conditions (\ref{1dcs10}), 
\beqar
D_a D_a &=& - C_n \nonumber\\
d_{abc} D_b D_c&=& -i \alpha_n  D_a\label{1dcs11}
\eeqar
Thus, even after the gauging $-iT_a \rightarrow -iT_a +A_a$, the derivatives obey the same embedding
conditions (\ref{1dcs10}) as before gauging \cite{KNP}.

The role of the gauge field is further clarified by the following identities which are easily proved.
The symbol for the product of $T_a$ with a matrix ${\hat A}$ is given by
\beqar
(T_a {\hat A}) 
&=& \left[ {nk\over \sqrt{2k(k+1)}} S_{a k^2+2k} - {1\over 2} \sum^k ~S_{a -i}  ~ R_{+i}
\right] A(g)\nonumber\\
({\hat A}T_a )&=& \left[  {nk\over \sqrt{2k(k+1)}} S_{a k^2+2k} + {1\over 2}
\sum^k ~S_{a +i}~R_{-i}
\right] A(g)
\label{1dcs12}
\eeqar
where $S_{a b} = 2~ \Tr (g^{-1} t_a g t_b )$.
If conditions 
(\ref{1dcs11}) are satisfied, then the symbol of $A_a$ is of the form $S_{ai}A_i$.
Equations (\ref{1dcs12}) then show that the commutator is equivalent to the derivative $R_i$, and
also that, upon gauging, we get $R_i \rightarrow R_i +A_i$. 

Now, 
if $K$ is a function on fuzzy ${\bf CP}^k$, by using the symbol, we can obtain its large $n$ expansion. We can think of $K$ as given in terms of a basis made of the identity matrix,
$(-iT_a)$  and sums of products of $(-iT)$'s, suitably orthonormalized. In the large $n$ expansion, the $T$'s get replaced by $S_{a k^2+2k} $ and $S_{ai}  R_i$, as in
equations (\ref{1dcs12}). This corresponds to expansion with only a background $U(1)$-field
for ${\bf CP}^k$ as in (\ref{1dcs3}). We can also expand the same matrix $K$ in terms of a basis  made of $D$'s rather than $(-iT)$'s. This will correspond to the ${\bf CP}^k$ having the gauge field $A_i$ in addition to the $U(1)$-field.

To carry out these expansions, we write $A_0$, $A_i$ in terms of $(N\times N)$-blocks.
In other words, we write the matrix elements of $A_i$, $i =0, 1, 2, ...$ as
$A_{i AB} = \la A \vert A_i \vert B\ra = \la \alpha ~a \vert A_i \vert \beta~ b\ra$,
$\alpha ,\beta = 1, 2, \cdots, N$, $a, b = 1, 2, \cdots, M$, corresponding to
the Hilbert space ${\cal H}_{\mathcal N}$ being a tensor product
${\cal H}_{N} \otimes {\cal H}_M$.
${\cal H}_{N}$ will carry an irreducible representation of $SU(k+1)$,
specifically the symmetric rank $n$ representation. 
In carrying out a large $n$ expansion, we will be rewriting the traces in terms
of integrals over (the star products of) the symbols as in (\ref{1dcs8}).
The symbols corresponding to a matrix in this direct product splitting is defined by
\beq
(A_i)_{ab} = \sum_{\alpha , \beta} {\cal D}^{(n)*}_{\alpha , n} (g)~ \la \alpha ~a \vert A_i \vert \beta~ b\ra ~{\cal D}^{(n)}_{\beta ,n}(g)
\label{1dcs12a}
\eeq
The symbol $(A_i )_{ab}$ is thus a matrix-valued function; it may be taken to be in the Lie algebra of $U(M)$. The trace of a matrix on ${\cal H}_{\cal N}$ then gets converted to the integral of the symbol over the manifold with a remaining trace over ${\cal H}_M$.

As mentioned at the beginning of this section, for a group coset manifold 
such as 
${\bf CP}^k = SU(k+1)/U(k)$, it is also possible to choose the background field to be nonabelian, valued in the Lie algebra of $U(k)$
\cite{KN}.
For such a case, the wave functions are of the form ${\cal D}^{(J)}_{A a}(g)$, where $J$ denotes a
representation of $SU(k+1)$, and we also have
a nontrivial representation of $SU(k)\in U(k)$ for the right translations of $g$, corresponding to the index $a$, in addition to the nontrivial action of $R_{k^2+2k}$.
The definition of the symbol then becomes
\beq
(A_i)_{ab} = \sum_{A, B} {\cal D}^{(J)*}_{A , a} (g)~ \la A \vert A_i \vert B \ra ~{\cal D}^{(J)}_{Bb}(g)
\label{1dcs12b}
\eeq
This definition, which was used in \cite{KN} and \cite{karabali}, is a little bit different
from (\ref{1dcs12a}). We shall use (\ref{1dcs12a}) in what follows.
The resulting expressions will involve functions which are also matrices in
${\cal H}_M$. If $U(M)$ contains $SU(k+1)$ as a subgroup, then, at this stage, 
we can introduce additional ${\cal D}$'s
using $\delta_{ab}  = {\cal D}^{K*}_{ac} {\cal D}^K_{bc}$, where these ${\cal D}^K$'s 
are in a suitable representation (or representations) of $SU(k+1)$
and reduce the product ${\cal D}^{(n)}_{\beta ,n}
{\cal D}^K_{bc}$ to get wave functions like ${\cal D}^{(J)}_{Bc}$. Results using (\ref{1dcs12b})
can thus be recovered from results obtained using (\ref{1dcs12a}).
Notice that a part of the remaining gauge group $U(M)$, namely $SU(k+1)$, is the isometry
group of the space, and so, this procedure can be viewed as gauging
the isometry group, the indices $a, b$ in (\ref{1dcs12a}) taking the place of the tangent space indices.

We now introduce the notation
$[D_a, D_b] = f_{abc} D_c
+F_{ab} \equiv \Omega_{ab}$, which is the definition of the field strength $F_{ab}$.
The function $K$ can be taken as the sum of terms of the form $K = K^{a_1 a_2 \cdots a_s} D_{a_1} D_{a_2} \cdots D_{a_s}$, where the coefficients $K^{a_1 a_2 \cdots a_s}$ 
can be taken to be symmetric in all indices. (Any antisymmetric pair may be reduced to a single $D$ and $F$; $F$
itself may be re-expanded in terms of $D$'s, to bring it to this form.)
Thus $K$ has the form
\beq
K = \int d\mu ~ e^{\bz \cdot D } ~K(z) \label{1dcs13}
\eeq
where
\beq
d\mu = \prod_a {dz_a d\bz_a \over \pi}~e^{-\bz \cdot z}, \hskip .5in
K(z) = K^{a_1 a_2 \cdots a_s} z_{a_1} z_{a_2}\cdots z_{a_s}
\label{1dcs14}
\eeq
We now want to express $K$ in an expansion around
a perturbed version of the $D$'s, namely, $D'_a$ where $A'_a = A_a +\delta A_a $. Clearly this can be achieved
by writing $D_a = D'_a - \delta A_a = D'_a - \delta D_a$ in the expression for $K$.
The change in $K$ is thus given by $D_a \rightarrow D_a -\delta D_a$.
Varying the expression (\ref{1dcs13}) and taking traces, we find
\beqar
\Tr~ \delta K &=& -\Tr~ (\delta D_a K^a )\nonumber\\
K^a&=& \int d\mu ~\bz^a ~e^{\bz \cdot D} ~K(z) = \int d\mu ~e^{\bz \cdot D}~ {\del K \over \del z_a}\label{1dcs15}
\eeqar
Again using (\ref{1dcs13}), we find
\beqar
[D_a ,K] &=& [D_a , D_b] K^b ~-~ {1\over 2} [[D_a, D_b],D_c]~K^{bc} ~+\cdots
\nonumber\\
&=&K^b [D_a , D_b] ~+~ {1\over 2} K^{bc} [[D_a, D_b],D_c]~+\cdots\label{1dcs16}\\
K^{bc}&=&  \int d\mu~ \bz^b \bz ^c ~e^{\bz \cdot D} ~K(z)\nonumber
\eeqar
We can define a quantity $N_{ab}$ by the formula \cite{nair2}
\beqar
N_{ac} \Omega_{cb}  &=& \delta_{ab} + {\mathbb X}_{ab} + i {\mathbb Y}_{ab} \nonumber\\
\Omega_{bc} N_{ca} &=&  \delta_{ba} + {\mathbb X}_{ba} + i {\mathbb Y}_{ba} 
\label{1dcs17}
\eeqar
where
\beq
{\mathbb X}_{ab} = {D_b D_a \over B_n}, \hskip .3in {\mathbb Y}_{ab}
= {1\over B_n}  \left( n +{\half} (k+1)\right) 
\left( d_{abc} D_c +i { \alpha_n\over 2} \delta_{ab}\right)
\label{1dcs18}
\eeq
$\Omega_{ab}$ does not have an inverse, but $N_{ab}$ serves the purpose when applied
on quantities which obey the embedding conditions (\ref{1dcs11}). 
$N_{ab}$ has the series solution
\beqar
N_{ab} &=& N_{0ab} - ({\mathbb R}_{ab} N_0 )_{ab}  - (N_0 F N_0 )_{ab} + \cdots\nonumber\\
N_{0ac} &=& {1\over B_n} \biggl[  f_{ack} D_k + {1\over 4} (k-1) \delta_{ac}\biggr]\nonumber\\
B_n &=& {n (n+k+1) \over 4} + {k^2 -1 \over 16}
\label{1dcs19}
\eeqar
We can now
simplify equation (\ref{1dcs15}) as
\beqar
\Tr~(\delta K) &=&-{1\over 2} \Tr \biggl[ \delta D_a N_{ab} (\Omega_{bc} K^c)
- \delta D_a (K^c \Omega_{bc}) N_{ba} \nonumber\\
&&\hskip 1.2in +
{1\over B_n} \bigl[\delta D_a (\Omega_{ab} K^b )
-\delta D_a (K^b \Omega_{ab}) 
\bigr] \biggr]\nonumber\\
&=&-{1\over 2} \Tr \Bigl[ \delta D_a N_{ab} [D_b, K] 
- [D_b, K] N_{ba} \delta D_a + {\cal O}(1/n^3)\Bigr]\label{1dcs20}
\eeqar
where we have used equation (\ref{1dcs16}). What is needed now is to write this in terms of symbols and simplify it. This is a fairly
long calculation, but eventually leads to the following result. 
More details of this calculation can be found
in \cite{nair2}.

The action of interest is the analog of (\ref{intro2}), with $U=1$. This is the one-dmensional Chern-Simons action $S= i \int dt~ \Tr (D_0)$. This is simplified as
\beq
i\int dt~ \Tr (D_0) \approx S_{*CS}~+~\cdots
\label{1dcs21}
\eeq
where $S_{CS}$ is the Chern-Simons $(2k+1)$-form on ${\bf CP}^k\times {\bf R}$.
This is defined by the variation
\beq
\delta S_{CS} =  {i^{k+1}\over (2\pi )^k k!} \int \tr (\delta A F^k)
\label{1dcs22}
\eeq
$S_{*CS}$ in (\ref{1dcs21}) is the star-version of the same action, i.e., $S_{CS}$ with star products connecting the various fields in it. The gauge potentials in the Chern-Simons action are given by
$a +A$, where $a$ is the background value corresponding to the symplectic form
and $A$ is the additional potential or gauge field fluctuation.
If we take the gradients of the fields to be small compared to their values, so that $\vert D^2 F\vert  \ll \vert F F\vert  $, for example, then the higher terms in the star product are negligible
and we can write \cite{nair2}
\beq
i\int dt~ \Tr (D_0) \approx S_{CS} (a+A) ~+~\cdots
\label{1dcs23}
\eeq
This is our basic result. Notice that only the combination $a+A$ appears in the action; thus the dependence on the specific choice of the ${\bf CP}^k$ background has disappeared.
Different choices of the background potentials correspond to different large $n$ limits.
Of course, the action is also sensitive to the dimension; the fact that we chose to expand
in terms of a $2k$-dimensional phase space shows up in equation
(\ref{1dcs23}). The large limits are thus parametrized by the gauge potentials and the choice of the dimension.

The action (\ref{1dcs23}) is also the action which arises for the lowest Landau level for
different gauge field backgrounds \cite{karabali}. Thus it is the action for the bulk dynamics 
for the quantum Hall system. We will now go back to the question of how this applies to gravity on a fuzzy space.

\section{A matrix version of gravity}

Gauge fields in the case of quantum Hall effect take values in the Lie algebra of $U(k)$.
From the matrix model point of view, where we write matrices in terms of the ${\cal H}_N
\otimes {\cal H}_M$ splitting, there is no obstruction to extending this to $SU(k+1)$ or
even any unitary group. (If we allow all possible types of fluctuations, it is a unitary group
$U(M)$ for some $M$, rather than $SU(M)$, that is relevant. This is also what is needed on noncommutative spaces.)
At this stage, it is worth recalling that ordinary Minkowski space may be considered as
the coset space $P/L$, where $P$ is the Poincar\'e group and $L$ is the Lorentz group. For the case of ${\bf CP}^k = SU(k+1)/U(k)$, the group $SU(k+1)$ is the analog of the Poincar\'e group
and the isotropy group $U(k)$ is he analog of the Lorentz group. The gauge fields we have introduced correspond to the gauging of these groups which are isometries of ${\bf CP}^k$,
implying that they should be interpreted as gravitational fields.
Based on the idea that the states which describe space itself should be treated exactly as the states for
matter are treated, we see that the simplification of the action in the previous section can be used.

There are, evidently, some missing ingredients which have to be taken care of before
this can be interpreted in terms of gravity.
First of all, the gauge fields are of the form $A^a_\mu dx^\mu (-it^a)$, which are one-forms on
${\bf CP}^k \times {\bf R}$ (because $dt$ is included in this expression) and the Lie algebra matrices form a basis for $U(k+1)$, or some other unitary group. There is, so far, no analog
of $e^0_\mu dx^\mu$ or $\omega^{0a}_\mu dx^\mu$, corresponding to the time-components of the frame field or spin connection.
Secondly, there is a dimensional mismatch since $U(k+1)$ gives $(k+1)^2$ one-form fields on a $(2k+1)$-dimensional space, where as we we need $(2k+1)(k+1)$ one-forms to describe gravity.
Thirdly, the matrix traces are naturally positive and lead to Euclidean signature for the tangent space. 
We shall now see how these issues can be addressed.

We start again with a finite-dimensional Hilbert space ${\cal H}$ describing the possible states
of a quantum system. 
At this stage we do not even have a notion of time, but just this Hilbert 
space ${\cal H}$
and, let us say, the
system under study is observed to be in a state given by a density matrix
$\rho_0$.
Any unitary transformation of $\rho_0$ is also an allowed density matrix, therefore, what constitutes a change in the system is a unitary transformation.
In other words, we can describe trajectories in ${\cal H}$ which correspond to sequences
of unitary transformations. The notion of evolution of the system arises when we are able to
compare the density matrix with the density matrix of another reference system.
Changes in the density matrix can be considered as a function of a sequence of changes
in the reference system and this is the meaning of time-evolution.
Let $t$ denote the parameter along the trajectory of  $\rho_0$ in ${\cal H}$. The motion along the trajectory is generated by some
hermitian operator $K$ defined by
$i {\dot U}  =  K~U$. (The parameter $t$ is not necessarily time, it is just the parameter along the trajectory. Later, after a large $N$ limit is taken, it may be identified with time.)
The density matrix $\rho \equiv U \rho_0 U^\dagger$ obeys the equation
\beq
i \del_0 \rho = i {\del \rho \over \del t} = [K, \rho ]
\label{grav1}
\eeq
This equation is essentially the definition of $K$. However, if $K$ is given, it can be taken as defining the evolution of $\rho$.
The action which leads to this equation is given by
\beq
S = i \int d t~ \Tr \bigl[\rho_0~ U^\dagger (\del_0 + A_0 ) U \bigr]
\equiv i \int dt~ \Tr \bigl[\rho_0~ U^\dagger D_0 U \bigr]
\label{grav2}
\eeq
where $A_\tau = i K$. 
Notice also that this action (\ref{grav2}) has a natural gauge invariance,
$U\rightarrow h~ U$, $ A_0 \rightarrow h A_0 h^\dagger - \del_0 h~ h^\dagger$.
The action (\ref{grav2}) is a function of $U$ and gives equation (\ref{grav1}) as the variational equation $\delta S =0$ for variations of $U$.

We now consider a separation of this quantum system into a 
part corresponding to the degrees of freedom of space and a part which describes all other, material, degrees of freedom, denoted as the subsystem $S$.
In other words,
${\cal H} = {\cal H}_{\cal N} \otimes {\cal H}_S$.
Correspondingly, a state in the Hilbert space may be represented as
$\vert A,r\ra$.
The labels $A, B,$ etc., pertain to the degrees of freedom of space
(or `environment')
and the labels $r, s,$ etc., describe the subsystem of interest. 
For the operator $D_0$, we introduce the splitting
\beq
\la A, r\vert D_0 \vert B, s\ra
=  \delta_{rs} ~\la A \vert D^{(e)}_{0}\vert B\ra ~+~ \la A, r\vert D^{(s)}_0
\vert B, s\ra
\label{grav3}
\eeq
The part of $D_0$ which is proportional to the identity in ${\cal H}_S$ is designated
as $D^{(e)}_0$ and the remainder as $D^{(s)}_0$. The latter includes effects
of coupling the subsystem of interest to the spatial degrees of freedom.
The density matrix also has a splitting of the form
\beq
\la A, r\vert \rho_0 \vert B, s\ra = \delta_{AB} ~\la r \vert \rho_0 \vert s\ra
\label{grav4}
\eeq
(This is not normalized, normalization will be taken care of later.)
The rank of the density matrix is a measure of how much of the Hilbert space is covered by
the chosen state of the system. If its rank is less than maximal for ${\cal H}_{\cal N}$, it would
mean that the dynamics does not cover all of space. This is why we choose it
to  be of maximal rank in ${\cal H}_{\cal N}$.

Our proposal for fuzzy gravity is then the following. We take the action (\ref{grav2})
as the action for the theory, including gravity, where $U$ and $D^{(e)}_0$ are regarded as 
quantities to be varied. $D^{(s)}_0$ is to be regarded as a given operator, specifying the subsystem of interest. 

The notion of continuous space emerges in the limit of the 
dimension of ${\cal H}_{\cal N}$ becoming large. One may regard 
${\cal H}_{\cal N}$ as arising from the quantization of some phase space ${\cal M}$,
with an appropriate symplectic form. The background fields on this phase space can be varied.
Thus it is possible to calculate the action, expanding $D^{(e)}_0$ in terms of 
the background gauge fields, in the limit of the dimension of ${\cal H}_{\cal N}$ becoming large. This is, of course, what we have done in the previous section. The best background to expand around is then given by the extremization of the action. 
Notice that the choice of spatial geometry in terms of these background fields arises
as a choice of which large ${\cal N}$ limit is best suited for the study of the
subsystem of interest. In this sense, the extremization with respect to these variables is
more like the choice of a thermodynamic state rather than a choice of dynamical trajectory.
The extremization with respect to $U$, with $D^{(s)}_0$ specifying the subsystem of interest
is the usual, dynamical, choice of a trajectory.

We now consider this idea in more detail. The action can be written as
\beq
S=  i \int dt ~ \sum_{A, r,s} (\rho_0)_{rs}~ \la A, s \vert ~U^\dagger
D_0 U ~\vert A, r\ra
\label{grav5}
\eeq
If the interaction between the material system and the spatial (gravitational) degrees of freedom is small, then, as a first approximation, we can look at the action for the
spatial degrees of freedom by itself. Thus
\beq
S \approx i \int dt~ \Tr ( D^{(e)}_0 ) 
\label{grav6}
\eeq
For expansion around ${\bf CP}^k$, we can simplify this, in the large $n$ limit and for slowly fields, to obtain
the Chern-Simons action $S_{CS}$ with the gauge fields being
$a +A$. Notice that this depends only on the full gauge field, the separation into
$a$ and the fluctuation $A$ is immaterial. In the following, we shall rename
the combination $a+A$ as $A$.
We now see the basic consequence of
our assumption that the spatial degrees of freedom in 
${\cal H}_{\cal N}$ should be treated in the same way as the dynamical degrees of freedom
in ${\cal H}_S$. It implies that the gravitational action, which determines the best geometry 
on which further analysis of the matter dynamics can be carried out, should be the Chern-Simons action \cite{zanelli}.
In particular, this will work only for odd dimensional spacetimes.
We also note that there are indications of Chern-Simons gravity in the context of
M-theory \cite{horava}; we expect that our approach is related to a matrix version of some of the
considerations in these references.

The gauge group which occurs in the CS action should be a unitary group, say, $U(M)$, since it arises
from the splitting
${\cal H}_{\mathcal N} = {\cal H}_{N} \otimes {\cal H}_M$,
with 
$A_{AB} = \la A \vert ~A ~\vert B\ra = \la \alpha ~a \vert ~ A ~\vert \beta~ b\ra$,
$\alpha ,\beta = 1, 2, \cdots, N$, $a, b = 1, 2, \cdots, M$.
While we interpret ${\cal H}_N$ as the quantization of ${\bf CP}^k$
with the symplectic form $\omega = -in \omega_K$, {\it a priori},
there is no restriction on $M$, we could choose any unitary group. However,
since ${\bf CP}^k = SU(k+1)/U(k)$,  the possible gauge field fluctuations which can be interpreted as gravity belong to $SU(k+1)$, or at best, $U(k+1)$. Therefore, we take
$M =k+1$. 

The question of the dimensional mismatch can be handled by using an idea similar to compactification. The simplest case for which this can be carried out is for $k=3$, corresponding to the group $U(4)$.
The Chern-Simons action is thus defined on a seven-dimensional space. We take this to be of the form $S^2 \times M^5$, where $M^5$ is some five-dimensional manifold.
Writing $U(4) \sim SU(4) \times U(1)$, the gauge field is taken to be of the form
$-i l \omega_K + {\cal F}$, where $\omega_K$ is the K\"ahler form of the two-sphere
$S^2$, $l$ is an integer and ${\cal F}$ belongs to the $SU(4)$ Lie algebra. The effective action is then given by the level $l$, five-dimensional Chern-Simons action with the gauge group $SU(4)$,
\beq
S = -i {l \over 24 \pi^2} \int \tr\left( A ~dA~ dA + {3\over 2} A^3~ dA +{3\over 5} A^5\right)
\label{grav7}
\eeq
Since $SU(4)$ is locally isomorphic to $O(6)$, we see that we have the correct set of gauge fields to describe Euclidean gravity in five dimensions.
In fact,  we can expand the gauge potential as
\beq
A = P^a ~e^a_\mu dx^\mu ~+~ \half  J^{ab} ~\omega^{ab}_\mu dx^\mu
\label{grav8}
\eeq
where $J_{ab}$ are the generators of $O(5) \subset O(6)$ and
$P_a$ are a basis for the complement of $\underline{O(5)}$ in $\underline{O(6)}$.
As a specific matrix representation, we take
$P_a = -(i/2) \gamma_a$, $J_{ab} = (1/4) [\gamma_a , \gamma_b ]$,
$\gamma$'s being the four-dimensional Dirac matrices.
In equation (\ref{grav8}), $e^a$ can be identified as the frame fields and $\omega^{ab}$ as
the spin connection.
The variation of the action (\ref{grav7}) can now be simplified as
\beq
\delta S =  - {l\over 128\pi^2} \int~\left[ \delta \omega^{ab} ~{\mathcal R}^{cd}~ ({\mathcal D} e)^e
+ {1\over 2} \delta e^a~ {\mathcal R}^{bc}~ {\mathcal R}^{de}\right]
\epsilon_{abcde}
\label{grav9}
\eeq
where
\beqar
({\mathcal D} e)^a &=& d e^a +\omega^{ac}~e^c\nonumber\\
{\mathcal R}^{ab} &=& R^{ab} ~-~ e^a ~e^b\label{grav10}\\
R^{ab}&=& d\omega^{ab} ~+~ \omega^{ac} ~\omega^{cb}\nonumber
\eeqar
We see that $R^{ab}$ is the Riemann tensor for the spin connection
and $({\mathcal D}e)^a$ is the torsion tensor.
The equations of motion for gravity with no matter field are then
\beqar
\epsilon_{abcde}~ {\mathcal R}^{cd}~ ({\mathcal D} e)^e &=& 0\nonumber\\
\epsilon_{abcde}~ {\mathcal R}^{bc}~ {\mathcal R}^{de}  &=& 0\label{grav11}
\eeqar

The solution to these equations, corresponding to empty space with no matter,
is thus given by
\beq
A = g^{-1} d g, \hskip .5in g \in O(6)
\label{grav12}
\eeq
This space is $O(6)/O(5) =S^5$ which is the Euclidean version of de Sitter space.
It is given in a basis where the
cosmological constant has been scaled out; it may be introduced
by the replacement
$e^a \rightarrow \sqrt{\Lambda} ~e^a$. The full solution of the CS action on the seven-dimensional space is thus $S^2 \times S^5$, where the $U(1)$ component has the nonzero value $-il \omega_K$ on the $S^2$ and the $O(6)$ fields have the nonzero value given by 
(\ref{grav12}) on the $S^5$.

There is also a neat reduction of this to four dimensions \cite{CMM}. Going back to the five-dimensional theory (\ref{grav7}), we can take $e^5_5 =1$, $\omega^{5a} =0$, $\omega^{ab}_5 =0$,
for $a, b = 1, ... , 4$. The fifth dimension is taken as a circle of, say, unit radius.
In this case, (\ref{grav7}) gives
\beqar
\delta S &=& - {l \over 32\pi} \int \left[ \delta \omega^{ab} e^c ( {\cal D} e )^d 
- \delta e^a e^b {\cal R}^{cd} \right] \epsilon_{abcd}\nonumber\\
&=&\delta \left[ {l \over 64\pi} \int \left( e^a e^b R^{cd} - {1\over 2}  e^a e^b e^c e^d \right)
\epsilon_{abcd} \right]
\label{grav13}
\eeqar
We see that the resulting four-dimensional action is
\beq
S = {l \Lambda \over 64\pi} \int \left( e^a e^b R^{cd} - {1\over 2}  e^a e^b e^c e^d \right)
\label{grav14}
\eeq
(We have also done the scaling $e^a \rightarrow \sqrt{\Lambda} ~e^a$.)
This action leads to the zero torsion condition and the vacuum Einstein equations with a cosmological constant. It can also be written in the Einstein-Hilbert form
\beq
S = {l \Lambda \over 16\pi} \int \sqrt{g} ~d^4x ~( R - 3 \Lambda ) 
\label{grav15}
\eeq
The choice of the conditions $e^5_5 =1$, $\omega^{5a} =0$, $\omega^{ab}_5 =0$
makes the theory into a nontopological theory.

The case of other dimensions can be treated similarly. For example, we could go back to the 
five-dimensinal case (\ref{grav7}) and introduce a compactification to three dimensions
by choosing $M^5 = S^2 \times M^3$ and picking a background value for $de^5$ as the 
K\"ahler two-form on the $S^2$. Upon integrating over this $S^2$, we will get a three-dimensional CS form with the gauge group $O(4)$. This can be interpreted as the action
for three-dimensional gravity as in \cite{CSgrav}.

One can also start in higher dimensions. For example, if we start with 11 dimensions,
which would correspond to a group $U(6)$, we can choose $M^{11}$ as ${\bf CP}^2
\times M^7$, where the ${\bf CP}^2$-space has a background which is $\underline{U(2)}$-valued. This gives a reduction to seven dimensions with $U(4)$ fields on it, and further reduction can be done as before. However, to get full-fledged gravity in 11 dimensions,
with $\underline{O(12)}$-valued gauge fields, we will need to start with the expansion of the CS one-form matrix action (\ref{grav6}) corresponding to a higher dimensional space
which has aunitary group which can accommodate $O(12)$.

We shall now comment briefly on the issue of Minkowski signature.
The action involves the trace over the matrix labels of $U(M)$. For the five-dimensional
theory, this corresponds to the spinor representation of $O(6)$, which is also the fundamental
representation of $SU(4)$. This is constructed in terms of the $\gamma$-matrices which obey
$\gamma_a \gamma_b + \gamma_b \gamma_a = 2\delta_{ab}$, $a,b =1,..., 5$. We can rewrite
the generators of $SU(4)$ in terms of a new set of $\gamma$-matrices, say,
${\tilde \gamma}_a = \gamma_a \gamma_4$, $a \neq 4$, ${\tilde \gamma}_0
= \gamma_4$. This leads to ${\tilde \gamma}_a {\tilde \gamma}_b +{\tilde \gamma}_b
{\tilde \gamma}_a = 2 \eta_{ab}$, where $\eta_{ab}$ is the Minkowski metric.
We can now expand the fields $A$ in terms of ${\tilde \gamma}_a$, ${\tilde \gamma}_{ab}$,
identifying the components as $O(5,1)$ gauge fields. This defines a particular
Minkowski continuation. While this can formally carry out a continuation of the tangent
frame indices, how the proper continuation can be done for the matter part is not yet clear.
We hope to address this question in more detail in a future publication.

\vskip .1in\noindent
I thank Dimitra Karabali for useful discussions.
This work
was supported in part by the National Science Foundation grant number
PHY-0244873 and by a PSC-CUNY grant.

%%%%%%%%%%%%%%%%%%%%%%%%%%%%%%%%%%%%%
\end{document}